\documentclass[english]{article}
\usepackage[T1]{fontenc}
\usepackage[latin9]{inputenc}
\usepackage{amsmath}
\usepackage{setspace}
\makeatletter
\newenvironment{lyxlist}[1]
{\begin{list}{}
{\settowidth{\labelwidth}{#1}
 \setlength{\leftmargin}{\labelwidth}
 \addtolength{\leftmargin}{\labelsep}
 }}
{\end{list}}

\makeatother

\usepackage{babel}

\begin{document}

\title{On the {}``Security analysis and improvements of arbitrated quantum
signature schemes'' }

\author{Song-Kong Chong, Yi-Ping Luo, Tzonelih Hwang%
\thanks{Corresponding Author%
}}
\maketitle
\begin{abstract}
Recently, Zou et al. {[}Phys. Rev. A 82, 042325 (2010){]} pointed
out that two arbitrated quantum signature (AQS) schemes are not secure,
because an arbitrator cannot arbitrate the dispute between two users
when a receiver repudiates the integrity of a signature. By using
a public board, they try to propose two AQS schemes to solve the problem.
This work shows that the same security problem may exist in their
schemes and also a malicious party can reveal the other party's secret
key without being detected by using the Trojan-horse attacks. Accordingly,
two basic properties of a quantum signature, i.e. unforgeability and
undeniability, may not be satisfied in their scheme.

\end{abstract}

\section{Introduction}

Quantum signature, which concerns about the authenticity and non-repudiation
of quantum states on an insecure quantum channel \cite{Zen2002,Cho2011_3},
is one of the most important researches in quantum cryptography. By
exploiting the principles of quantum mechanics, e.g., no-cloning theory
and measurement uncertainty, quantum signature can provide unconditional
security. Two basic properties are required in a quantum signature
\cite{Zen2002} :
\begin{enumerate}
\item Unforgeability: Neither the signature verifier nor an attacker can
forge a signature, or change or attach the content of a signature.
The signature should not be reproduced by any other person.
\item Undeniability: A signatory, Alice, who has sent the signature to the
verifier, Bob, cannot later deny having signed a signature. Moreover,
the verifier Bob cannot deny the receipt of the signature. 
\end{enumerate}
Quantum signature was first investigated by Gottesman and Chuang \cite{Got2001}.
After that, a variety of quantum signature schemes have been proposed
\cite{Zen2002,Cho2011_3,Yan2010,Cho2011,Cur2008,Zen2008,Li2009,Yan2010-1,Wan2005_3,Lee2004,Wan2006_3,Wen2006,Wen2007_2}.
Zeng et al. \cite{Zen2002} proposed an arbitrated quantum signature
(AQS) scheme based on the correlation of GHZ states and quantum one-time
pads. However, Curty et al. \cite{Cur2008} pointed out that \cite{Zen2002}
is not clearly described and the security statements claimed by the
authors are incorrect. In the reply comment \cite{Zen2008}, Zeng
gave a more detailed presentation and proof to their original AQS
scheme \cite{Zen2002}. To improve the transmission efficiency and
to reduce the implementation complexity of \cite{Zen2002,Zen2008},
Li et al. \cite{Li2009} proposed an AQS scheme using Bell states
and claimed that their improvements can preserve the merits in the
original scheme \cite{Zen2002,Zen2008}. 

In an AQS scheme, an arbitrator plays a crucial role. When a dispute
arises between the users, the arbitrator should be able to arbitrate
the dispute. The arbitrator should be able to solve a dispute when
a receiver, Bob, repudiates the receipt of the signature, or in particular,
the receiver repudiates the integrality of the signature, i.e., Bob
admits receiving a signature but denies the correctness of the signature.
The dispute of the latter one implies the following three cases \cite{Zou2010}:
\begin{description}
\item [{\textmd{(1)}}] Bob told a lie;
\item [{\textmd{(2)}}] The signatory Alice sent incorrect information to
Bob;
\item [{\textmd{(3)}}] An eavesdropper Eve disturbed the communications.
\end{description}
Since the arbitrator in \cite{Zen2002,Zen2008,Li2009} cannot solve
the dispute when Bob claims that the verification of a signature is
not successful, Zou et al. \cite{Zou2010} considered that these schemes
are not valid because the security requirement of a quantum signature,
i.e., the undeniability, is not satisfied. 

By using a public board, Zou et al. also proposed two AQS schemes
to solve the problem. However, this study will point out that the
same security problem may exist in their schemes. That is, when Bob
announces that the verification is not successful, the arbitrator
may not be able to distinguish which case described above has happened.
Besides, this study also tries to investigate if a malicious signer,
Alice, can reveal Bob's secret key without being detected by performing
the Trojan-horse attacks \cite{Cai2006,Deng2005_3}. 

The rest of this paper is organized as follows. Section 2 reviews
Zou et al.'s schemes. Section 3 shows the problems with the schemes.
Finally, Section 4 concludes the result.

\section{Review of Zou et al.'s schemes }

Zou et al.'s AQS schemes \cite{Zou2010} are briefly explained in
the following scenario. Alice, the message signatory, would like to
sign a quantum message $\left|P\right\rangle $ to a signature verifier,
Bob, via the assistance of an arbitrator, Trent. Suppose that Alice
and Bob share a secret key $K\in\left\{ 0,1\right\} ^{*}$, and the
quantum message to be signed is $\left|P\right\rangle =\left|P_{1}\right\rangle \otimes\left|P_{2}\right\rangle \otimes...\otimes\left|P_{n}\right\rangle $,
where $\left|K\right|\geq2n$, $\left|P_{i}\right\rangle =\alpha_{i}\left|0\right\rangle +\beta_{i}\left|1\right\rangle $,
and $1\leq i\leq n$. In order to protect the quantum message, the
quantum one-time-pad encryption $E_{K}$ \cite{Boy2003} and the unitary
transformation $M_{K}$ used in the schemes are defined as follows.

\begin{equation}
E_{K}\left(\left|P\right\rangle \right)=\overset{n}{\underset{i=1}{\bigotimes}}\sigma_{x}^{K_{2i-1}}\sigma_{z}^{K_{2i}}\left|P_{i}\right\rangle ,\end{equation}

\begin{equation}
M_{K}\left(\left|P\right\rangle \right)=\overset{n}{\underset{i=1}{\bigotimes}}\sigma_{x}^{K_{i}}\sigma_{z}^{K_{i\oplus1}}\left|P_{i}\right\rangle ,\end{equation}

\noindent where $\left|P_{i}\right\rangle $ and $K_{i}$ denote the
$i$th bit of $\left|P\right\rangle $ and $K$, $\sigma_{x}$ and
$\sigma_{z}$ are the Pauli matrices, respectively.

To prevent the integrality of a signature from being disavowed by
Bob, Zou et al. proposed two AQS schemes: the AQS scheme using Bell
states and the AQS without using entangled states, respectively. Their
schemes are described as follows.

\subsection{Scheme 1: the AQS scheme using Bell states}

Suppose that Alice wants to sign an $n$-bit quantum message $\left|P\right\rangle $
to Bob. In order to perform the signature, three copies of $\left|P\right\rangle $
are necessary. The scheme proceeds as follows:
\begin{lyxlist}{00.00.0000}
\item [{\textbf{Initializing}}] \noindent \textbf{phase:}\end{lyxlist}
\begin{description}
\item [{Step$\;\mathbf{\mathit{I}}\mathbf{1}.$}] The arbitrator Trent
shares the secret keys $K_{A},K_{B}$ with Alice and Bob respectively
through some unconditionally secure quantum key distribution protocols.
\item [{Step$\; I\mathbf{2}.$}] Alice generates $n$ Bell states, $\left|\psi_{i}\right\rangle =\frac{1}{\sqrt{2}}\left(\left|00\right\rangle _{AB}+\left|11\right\rangle _{AB}\right)$,
where $1\leq i\leq n$, and the subscripts $A$ and $B$ denote the
$1^{st}$ and the $2^{nd}$ particles of that Bell state, respectively.
After that, Alice sends all $B$ particles to Bob in a secure and
authenticated way \cite{Cur2002,Bar2002}.\end{description}
\begin{lyxlist}{00.00.0000}
\item [{\textbf{Signing$\;\:$phase}}]~\end{lyxlist}
\begin{description}
\item [{Step$\; S\mathbf{1}.$}] Alice chooses a random number $r\in\left\{ 0,1\right\} ^{2n}$
to encrypt all $\left|P\right\rangle $'s, i.e., $\left|P'\right\rangle =E_{r}\left(\left|P\right\rangle \right)$.
\item [{Step$\; S\mathbf{2}.$}] Alice generates $\left|S_{A}\right\rangle =E_{K_{A}}\left(\left|P'\right\rangle \right)$.
\item [{Step$\; S\mathbf{3}.$}] Alice combines each $\left|P_{i}'\right\rangle $
and the Bell state to obtain a three-particle entangled state, \textbf{\footnotesize \[
\left|\phi_{i}\right\rangle =\left|P_{i}'\right\rangle \otimes\left|\psi_{i}\right\rangle _{AB}=\frac{1}{2}\left[\begin{array}{c}
\left|\Phi_{PA}^{+}\right\rangle _{i}\left(\alpha_{i}^{'}\left|0\right\rangle +\beta_{i}^{'}\left|1\right\rangle \right)_{B}+\left|\Phi_{PA}^{-}\right\rangle _{i}\left(\alpha_{i}^{'}\left|0\right\rangle -\beta_{i}^{'}\left|1\right\rangle \right)_{B}+\\
\left|\Psi_{PA}^{+}\right\rangle _{i}\left(\alpha_{i}^{'}\left|1\right\rangle +\beta_{i}^{'}\left|0\right\rangle \right)_{B}+\left|\Psi_{PA}^{-}\right\rangle _{i}\left(\alpha_{i}^{'}\left|1\right\rangle -\beta_{i}^{'}\left|0\right\rangle \right)_{B}\end{array}\right],\]
}where $\left|\Phi_{PA}^{+}\right\rangle ,\left|\Phi_{PA}^{-}\right\rangle ,\left|\Psi_{PA}^{+}\right\rangle $,
and $\left|\Psi_{PA}^{-}\right\rangle $ are the four Bell states
\cite{Kwi1995}.
\item [{Step$\; S\mathbf{4}.$}] Alice performs a Bell-measurement on each
$\left|\phi_{i}\right\rangle $ and obtains the measurement results
$\left|M_{A}\right\rangle =\left(\left|M_{A}^{1}\right\rangle ,\left|M_{A}^{2}\right\rangle ,\ldots,\left|M_{A}^{n}\right\rangle \right)$,
where $\left|M_{A}^{i}\right\rangle \in\left\{ \left|\Phi_{PA}^{+}\right\rangle _{i},\left|\Phi_{PA}^{-}\right\rangle _{i},\left|\Psi_{PA}^{+}\right\rangle _{i},\left|\Psi_{PA}^{-}\right\rangle _{i}\right\} $,
and $1\leq i\leq n$. 
\item [{Step$\; S\mathbf{5}.$}] Alice sends $\left|S\right\rangle =\left(\left|P'\right\rangle ,\left|S_{A}\right\rangle ,\left|M_{A}\right\rangle \right)$
to Bob.\end{description}
\begin{lyxlist}{00.00.0000}
\item [{\textbf{Verification}}] \textbf{phase:}\end{lyxlist}
\begin{description}
\item [{Step$\;\mathbf{\mathit{V}1}.$}] Bob encrypts $\left|P'\right\rangle $
and $\left|S_{A}\right\rangle $ with $K_{B}$ and sends the quantum
ciphertext $\left|Y_{B}\right\rangle =E_{K_{B}}(\left|P'\right\rangle ,$
$\left|S_{A}\right\rangle )$ to Trent.
\item [{Step$\;\mathbf{\mathit{V}2}.$}] Trent decrypts $\left|Y_{B}\right\rangle $
with $K_{B}$ and obtains $\left|P'\right\rangle $ and $\left|S_{A}\right\rangle $.
Then he encrypts $\left|P'\right\rangle $ with $K_{A}$ and gets
$\left|S_{T}\right\rangle $. If $\left|S_{T}\right\rangle =\left|S_{A}\right\rangle $
\cite{Li2009,Buh2001}, Trent sets the verification parameter $V=1$;
otherwise, $V=0.$
\item [{Step$\; V\mathbf{3}.$}] Trent recovers $\left|P'\right\rangle $
from $\left|S_{T}\right\rangle $. Then he encrypts $\left|P'\right\rangle ,\left|S_{A}\right\rangle $
and $V$ with $K_{B}$ and sends the quantum ciphertext $\left|Y_{T}\right\rangle =E_{K_{B}}\left(\left|P'\right\rangle ,\left|S_{A}\right\rangle ,V\right)$
to Bob.
\item [{Step$\; V\mathbf{4}.$}] Bob decrypts $\left|Y_{T}\right\rangle $
and gets $\left|P'\right\rangle ,\left|S_{A}\right\rangle $, and
$V$. If $V=0$, Bob rejects the signature; otherwise, Bob continues
to the next step.
\item [{Step$\; V\mathbf{5}.$}] Based on Alice's measurement results $M_{A}$,
Bob can obtain $\left|P'_{B}\right\rangle $ from the $B$ particles
received from the Step $I\mathbf{2}$ according to the principle of
teleportation \cite{Li2009}. Then he compares $\left|P_{B}'\right\rangle $
with $\left|P'\right\rangle $. If $\left|P_{B}'\right\rangle =\left|P'\right\rangle $,
Bob informs Alice to publish $r$ and proceeds to the next step; otherwise,
he rejects the signature.
\item [{Step$\; V\mathbf{6}.$}] Alice publishes $r$ on the public board.
\item [{Step$\; V\mathbf{7.}$}] Bob recovers $\left|P\right\rangle $
from $\left|P'\right\rangle $ by $r$ and holds $\left(\left|S_{A}\right\rangle ,r\right)$
as Alice's signature for the quantum message $\left|P\right\rangle $.
\end{description}

\subsection{Scheme 2: the AQS scheme without using entangled states}

Since the preparation, distribution, and storing of quantum entangled
states are not easily implemented with today's technologies, Zou et
al. also proposed an AQS scheme without using entangled states (Scheme
2) in the signing phase and the verifying phase. In order to prevent
a signature from being disavowed by Bob, a public board is also used
in the proposed scheme. The scheme is described as follows.
\begin{lyxlist}{00.00.0000}
\item [{\textbf{Initializing}}] \noindent \textbf{phase:}\end{lyxlist}
\begin{description}
\item [{Step$\;\mathbf{\mathit{I}}\mathbf{1'}.$}] The arbitrator Trent
shares the secret keys $K_{AT},K_{BT}$ with Alice and Bob respectively
through some unconditionally secure quantum key distribution protocols.
Similarly, Alice shares a secret key, $K_{AB}$, with Bob.\end{description}
\begin{lyxlist}{00.00.0000}
\item [{\textbf{Signing$\;\:$phase:}}]~\end{lyxlist}
\begin{description}
\item [{Step$\; S\mathbf{1}'.$}] Alice chooses a random number $r\in\left\{ 0,1\right\} ^{2n}$
and then computes $\left|P'\right\rangle =E_{r}\left(\left|P\right\rangle \right)$
and $\left|R_{AB}\right\rangle =M_{K_{AB}}\left(\left|P'\right\rangle \right)$,
where $\left|P\right\rangle $ is as defined before.
\item [{Step$\; S\mathbf{2}'.$}] Alice generates $\left|S_{A}\right\rangle =E_{K_{AT}}\left(\left|P'\right\rangle \right)$.
\item [{Step$\; S\mathbf{3}'.$}] Alice generates $\left|S\right\rangle =E_{K_{AB}}\left(\left|P'\right\rangle ,\left|R_{AB}\right\rangle ,\left|S_{A}\right\rangle \right)$
as her signature and then sends it to Bob.\end{description}
\begin{lyxlist}{00.00.0000}
\item [{\textbf{Verification}}] \textbf{phase:}\end{lyxlist}
\begin{description}
\item [{Step$\;\mathbf{\mathit{V}1'}.$}] Bob decrypts $\left|S\right\rangle $
with $K_{AB}$ and obtains $\left|P'\right\rangle ,\left|R_{AB}\right\rangle $
and $\left|S_{A}\right\rangle $. Then he generates $\left|Y_{B}\right\rangle =E_{K_{BT}}\left(\left|P'\right\rangle ,\left|S_{A}\right\rangle \right)$
and sends it to Trent.
\item [{Step$\;\mathbf{\mathit{V}2'}.$}] Trent decrypts $\left|Y_{B}\right\rangle $
with $K_{BT}$ and obtains $\left|P'\right\rangle $ and $\left|S_{A}\right\rangle $. 
\item [{Step$\;\mathbf{\mathit{V}3'}.$}] Trent decrypts $\left|S_{A}\right\rangle $
with $K_{AT}$ to obtain $\left|P'_{T}\right\rangle $. If $\left|P'_{T}\right\rangle =\left|P'\right\rangle $,
he sets the verification parameter $V_{T}=1$; otherwise, $V_{T}=0$.
Then Trent announces $V_{T}$ on the public board. If $V_{T}=1$,
he regenerates $\left|Y_{B}\right\rangle $ and sends it back to Bob.
\item [{Step$\;\mathbf{\mathit{V}4'}.$}] If $V_{T}=0$, Bob rejects the
signature. For otherwise, he decrypts $\left|Y_{B}\right\rangle $
with $K_{BT}$ to obtain $\left|P'\right\rangle $ and $\left|S_{A}\right\rangle $.
Then he computes $\left|P_{B}'\right\rangle =M_{K_{AB}}^{-1}\left(\left|R_{AB}\right\rangle \right)$
and compares it with $\left|P'\right\rangle $. If $\left|P_{B}'\right\rangle =\left|P'\right\rangle $,
he sets the verification parameter $V_{B}=1$; otherwise, $V_{B}=0$.
Bob announces $V_{B}$ on the public board.
\item [{Step$\;\mathbf{\mathit{V}5'}.$}] If $V_{B}=0$, Alice and Trent
abort the scheme; otherwise, Alice announces $r$ on the public board.
\item [{Step$\;\mathbf{\mathit{V}6'}.$}] Bob recovers $\left|P\right\rangle $
from $\left|P'\right\rangle $ by $r$ and holds $\left(\left|S_{A}\right\rangle ,r\right)$
as Alice's signature for the quantum message $\left|P\right\rangle $.
\end{description}

\section{Problems to be discussed}

This section tries to investigate problems that could arise on Zou
et al.'s schemes if precautions are not taken. We first discuss the
deniable dilemma. Then, we investigate the Trojan-horse attacks against
the schemes.

\subsection{The deniable dilemma}

In Zou et al.'s schemes, the signatory Alice uses a random number
$r$ to protect the quantum message $\left|P\right\rangle $ (i.e.,
$\left|P'\right\rangle =E_{r}\left(\left|P\right\rangle \right)$)
before signing it. After the arbitrator Trent's verification, Bob
recovers $\left|P'_{B}\right\rangle $ and compares it with $\left|P'\right\rangle $.
Once Bob informs Alice that $\left|P_{B}'\right\rangle =\left|P'\right\rangle $,
Alice will publish $r$ on the public board, which is assumed to be
free from being blocked, injected or alternated. Finally, Bob recovers
$\left|P\right\rangle $ from $\left|P'\right\rangle $ by $r$ and
retains $\left(\left|S_{A}\right\rangle ,r\right)$ as Alice's signature.

It appears that if Bob informs Alice to publish $r$ on the public
board, then he cannot disavow the integrality of the signature. Accordingly,
Zou et al. considered that the use of the public board can prevent
the denial attack from Bob. However, if Bob claims that $\left|P_{B}'\right\rangle \neq\left|P'\right\rangle $
in Step V5 (or Step V4' in Scheme 2), Trent cannot arbitrate the dispute
between Alice and Bob because the following three cases are possible.
(This is particularly serious, if the signature scenario occurs in
an electronic block market, where Alice is a buyer and Bob, a block
company.)
\begin{enumerate}
\item Bob told a lie: In this case, Bob decides to forgo the recovery of
the message $\left|P\right\rangle $ due to some unknown reasons;
\item Alice sent incorrect information to Bob: In Step S3 of Scheme 1, Alice
deliberately generated $\left|\phi_{i}\right\rangle $ by another
message $\left|\hat{P}_{i}'\right\rangle $ with $\left|\hat{P}_{i}'\right\rangle \neq\left|P_{i}'\right\rangle $
or generated $\left|S\right\rangle =\left(\left|P'\right\rangle ,\left|S_{A}\right\rangle ,\left|M'_{A}\right\rangle \right)$
with $\left|M'_{A}\right\rangle \neq\left|M_{A}\right\rangle $ in
Step S5. In Scheme 2, Alice intentionally sent $\left|S\right\rangle =E_{K_{AB}}\left(\left|P'\right\rangle ,\left|\hat{R}_{AB}\right\rangle ,\left|S_{A}\right\rangle \right)$
with $\left|\hat{R}_{AB}\right\rangle \neq\left|R_{AB}\right\rangle $
to Bob in Step S3';
\item Eve disturbed the communication.
\end{enumerate}
Apparently, when Bob claims that $\left|P_{B}'\right\rangle \neq\left|P'\right\rangle $,
Trent cannot solve the dispute. Furthermore, as also pointed out in
\cite{Zou2010}, the signer, Alice, is able to publish an arbitrary
$r'\left(\neq r\right)$ in her favor without been verified, which
is obviously against the requirement of a signature scheme.

\subsection{The Trojan-horse attack}

In Zou et al.'s schemes, there are two transmissions of the same quantum
signals, i.e. first from Alice to Bob, and then from Bob to the arbitrator.
Therefore, the malicious Alice can reveal Bob's secret key without
being detected by performing the Trojan-horse attacks \cite{Cai2006,Deng2005_3}.
Similar to \cite{Cho2011}, there are two attack strategies in the
Trojan-horse attacks: the invisible photon eavesdropping \cite{Cai2006}
and the delay photon eavesdropping \cite{Deng2005_3}. The following
will discuss the invisible photon eavesdropping (IPE) on Zou et al.'s
schemes and show that Alice can obtain Bob's secret key without being
detected. Note that, Alice can also use the delay photon eavesdropping
to reveal Bob's secret key in the same way.

In Scheme 1, in order to reveal Bob's secret key $K_{B}$, Alice can
perform the IPE attack on the communications in Step $S5$ and Step
$V1$ as follows:
\begin{description}
\item [{Step$\; S\mathbf{5a}.$}] Alice first prepares a set of eavesdropping
states, $D^{i}\in\left\{ \frac{1}{\sqrt{2}}\left(\left|00\right\rangle +\right.\right.$
$\left.\left.\left|11\right\rangle \right)_{d_{1}^{i}d_{2}^{i}}\right\} $,
as invisible photons, where the subscripts $d_{1}^{i}$ and $d_{2}^{i}$
represent the $1^{st}$ and $2^{nd}$ photons in $D^{i}$, $1\leq i\leq n$.
For each state in $\left|P'\right\rangle $ (or $\left|S_{A}\right\rangle $),
Alice inserts $d_{1}^{i}$ as an invisible photon to that state and
forms a new sequence $\left|P'\right\rangle ^{d_{1}}$ ($\left|S_{A}\right\rangle ^{d_{1}}$).
Then Alice sends $\left|S\right\rangle ^{d_{1}}=\left(\left|P'\right\rangle ^{d_{1}},\left|S_{A}\right\rangle ,\left|M_{A}\right\rangle \right)$
to Bob.
\item [{Step$\;\mathbf{\mathit{V}1a}.$}] Bob encrypts $\left|P'\right\rangle ^{d_{1}}$
and $\left|S_{A}\right\rangle $ with $K_{B}$ and sends the quantum
ciphertext $\left|Y_{B}\right\rangle ^{d_{1'}}=E_{K_{B}}(\left|P'\right\rangle ^{d_{1}},$
$\left|S_{A}\right\rangle )$ to Trent. Before Trent receives the
quantum ciphertext $\left|Y_{B}\right\rangle ^{d{}_{1'}}$, Alice
captures $d_{1'}$ from $\left|Y_{B}\right\rangle ^{d_{1'}}$ and
measures $d_{1'}d_{2}$ together with the Bell measurement. According
to the measuring result of $d_{1'}^{i}d_{2}^{i}$, Alice can obtain
Bob's secret key $K_{B}^{2i-1,2i}$. 
\end{description}
Note that, Alice can also use the similar ways mentioned above to
obtain Bob's secret key $K_{BT}$ in Scheme 2. Since both Scheme 1
and 2 are insecure to the Trojan-horse attacks, Bob can deny having
verified a signature. Therefore, the basic properties of a quantum
signature, i.e. unforgeability and undeniability, are not satisfied
in their schemes.

\section{Conclusions}

This study has pointed out two security flaws in Zou et al.'s AQS
schemes, in which the arbitrator cannot arbitrate the dispute between
Alice and Bob when Bob claims failure in his verification. Besides,
a malicious signer can obtain verifier's secret key by performing
the Trojan-horse attacks. How to improve their AQS schemes to avoid
the problems mentioned in this paper will be an interesting future
research.

\section*{Acknowledgment}

This research is supported partially by National Science Council,
Taiwan, R.O.C., under the Contract No. NSC 98-2221-E006-097-MY3.

\bibliographystyle{IEEEtran}
\addcontentsline{toc}{section}{\refname}\bibliography{YIPING,ISLAB}

\end{document}